\documentclass{webofc}
\usepackage[varg]{txfonts}   
\pdfoutput=1
\usepackage{pstricks}
\usepackage{color}
\usepackage{amssymb,amsmath,bbm}
\usepackage{epsf}
\usepackage{epsfig}
\usepackage{afterpage}
\usepackage{longtable}
\usepackage{latexsym,mathrsfs,dsfont}
\usepackage{graphics}
\usepackage{url}
\usepackage{paralist}
\usepackage{bbold}

\newcommand{\vcb}{|V_{cb}|}
\newcommand{\vtd}{|V_{td}|}
\newcommand{\vub}{|V_{ub}|}
\newcommand{\vts}{|V_{ts}|}
\newcommand{\vus}{|V_{us}|}

\newcommand{\IM}{\rm{Im}}

\def\epe{\varepsilon'/\varepsilon}
\newcommand{\be}{\begin{equation}}
\newcommand{\ee}{\end{equation}}
\newcommand{\bi}{\begin{itemize}}
\newcommand{\ei}{\end{itemize}}

\def\kpn{K^+\rightarrow\pi^+\nu\bar\nu}

\def\klpn{K_{L}\rightarrow\pi^0\nu\bar\nu}

\def\ksm{K_S\to\mu^+\mu^-}

\usepackage{tikz}

\woctitle{Hunting New Animalcula}

\begin{document}

\begin{flushright}
        {AJB-24-2}
\end{flushright}

\vspace{7mm}

\title{\boldmath Hunting New Animalcula with Rare K and B Decays   \unboldmath}
%
%
\author{ Andrzej~J.~Buras\fnsep\thanks{Talk given at QCD@Work 2024, Trani, June 2024.} 
}
\institute{TUM Institute for Advanced Study, Lichtenbergstr. 2a, D-85748 Garching, Germany,\\
Physik Department, TUM School of Natural Sciences, TU M\"unchen, James-Franck-Stra{\ss}e,\\ D-85748 Garching, Germany}

\abstract{%
  We summarize the recent strategy for an efficient hunting of  new animalcula
  with the help of rare K and B decays that avoids the use of the $\vcb$ and $\vub$ parameters that are subject to tensions between their determinations from
  inclusive and exclusive decays. In particular we update the values
  of the $\vcb$-independent ratios of various K and B decay branching ratios predicted by the Standard Model. We also stress the usefulness of the $\vcb-\gamma$ plots in the search for new physics. We select {\em the magnificant seven} among rare K and B decays
  that should play a leading role in the search for new physics due
to their theoretical cleanness: $B^+\to K^+(K^*)\nu\bar\nu$ and $K^+\to\pi^+\nu\bar\nu$ measured recently by Belle II and NA62, respectively, $\klpn$ investigated by KOTO  and also  $B_{s,d}\to\mu^+\mu^-$ and $K_S\to\mu^+\mu^-$ measured by the LHCb, CMS and ATLAS experiments at CERN.}
\maketitle
\section{Introduction}
\label{intro}
The year 1676 was a very important year for the humanity. In this year 
Antoni van Leeuvenhoek (1632-1723) discovered the empire of bacteria. 
He called these small creatures {\it animalcula} (small animals). This 
discovery was a mile stone in our civilization  for at least two reasons: 
\begin{itemize}
\item
 He discovered invisible to us creatures which over thousands of years 
 were systematcally killing the humans, often responsible for millions 
 of death in one year. While Antoni van Leeuvanhoek did not know that 
 bacteria could be dangerous for humans, his followers like L. Pasteur (1822-1895),
 Robert Koch (1843-1910) and other {\it microbe hunters} not only realized 
 the danger coming from this tiny creatures but also developed weapons against
 this empire. 
\item
 He was the first human who looked at short distance scales invisible to 
 us, discovering thereby
 a new {\it underground world}. At that time researchers 
 looked mainly at large distances, discovering new planets and finding 
 laws, like Kepler laws, that Izaak Newton was able to derive from his 
 mechanics.
 \end{itemize}

While van Leeuvanhoek could reach the resolution down to roughly 
$10^{-6}$m, over the last  350 years this resolution could be improved 
by many orders of magnitude. On the way down to shortest distance 
scales scientists discovered {\it nanouniverse} ($10^{-9}$m), 
{\it femtouniverse}   ($10^{-15}$m) relevant for nuclear physics 
and low energy elementary particle physics and finally 
{\it attouniverse} ($10^{-18}$m)
that is the territory of contemporary high energy elementary particle physics.

In this decade and the coming decades we will be able to improve the resolution of  the short distance scales by at least
 an order of magnitude, extending the picture of fundamental physics 
down to scales $5\cdot 10^{-20}$m with the help of the high energy processes at the Large Hadron Collider (LHC). Further resolution, 
down to scales as short as $10^{-21}$m ({\it zeptouniverse}) or even shorter scales, 
 should be possible with the help of 
high precision experiments in which flavour violating processes played  a 
prominent role for decades \cite{Buras:2023qaf}. These notes deal with the latter route to the short 
distance scales which is an indirect route based solely on quantum
fluctuations.

The main goal of these efforts is not only the curiosity of whether  new animalcula beyond the ones of  the Standard Model (SM), already discovered, exist. The main motivation is the hope that finding them will help us to answer
a number of very important questions that the SM cannot answer. They are well known and will not be listed here.

The main strategy is at first sight very simple. One calculates various observables like
branching ratios of various decays of mesons and leptons within the SM and compares them with the experimental data. Finding the departures
from SM predictions, called these days {\em anomalies}, signals the
presence of new animalcula that through quantum fluctuations affect  SM
predictions. In order to identify the nature of these animalcula it is
crucial to measure many of these anomalies as precisely as possible.

In a given {\em  measurement } of an anomaly both experimentalists and theorists are
involved simply because in order to find departures from SM predictions
one needs both precise experimental data and precise theory.
Once these two requirements are satisfied, there are several routes one
can follow. The most common one found in the literature are global fits
in concrete new physics (NP) models in which these anomalies could be explained.

Another route is to study first the patterns of anomalies observed in 
the data and compare them  with the patterns of deviation from  SM
predictions in a given NP scenario. Such patterns, that expose suppressions and enhancements
of various observables relative to SM predictions, can be considered as DNAs of the animalcula hunted by us.
In particular the
correlations between various enhancements and suppressions can rule out
some NP scenarios already before any global fit is performed. This
is the strategy proposed by Jennifer Girrbach-Noe and myself in
2013 \cite{Buras:2013ooa}. It has been documented in several subsequent
papers, in particular in my book \cite{Buras:2020xsm} and recently in
\cite{Buras:2024mhy} with many colourful plots. Therefore I will not discuss
it here.

The present notes deal with a strategy for obtaining most precise SM predictions
for rare K and B meson decays to date, a very importat ingredient in the
indirect search for NP. It has been developed this time in collaboration with Elena Venturini \cite{Buras:2021nns,Buras:2022wpw}.

The strategy in question, to be called {\bf BV-strategy} for simplicity,  has been motivated by the problems in the determination of the 
CKM parameters that play a very important role in SM predictions
for rare K and B decays. Their values are usually obtained from global fits
dominated by  UTfitter \cite{UTfit:2022hsi} and CKMfitter  \cite{Charles:2004jd}.
However as stressed in \cite{Buras:2021nns,Buras:2022wpw} and later in 
\cite{Buras:2022qip} this determination is presently problematic for  
the following reasons:
\begin{itemize}
\item
  In a global fit which contains processes that could be infected by NP the resulting
  CKM parameters are also infected by it and consequently the resulting
  branching ratios cannot be considered as genuine SM predictions.
  Consequently the resulting deviations from  SM predictions obtained in
  this manner (the pulls) are not the deviations one would find if
  the CKM parameters were not infected by NP.
\item
  Tensions in the determinations of $\vcb$ and $\vub$ from inclusive and exclusive tree-level decays \cite{Bordone:2021oof,Finauri:2023kte,FlavourLatticeAveragingGroupFLAG:2024oxs}. Using these results lowers the precision with which CKM parameters can be
  determined and consequently also lowers the precision of SM predictions for
  many observabies as illustrated soon. Therefore 
  the inclusion of these determinations in the fit should be avoided until theorists
  agree what the values of  $\vcb$ and $\vub$ are. 
\item
  Hadronic uncertainties in some observables included in the fit are much larger than in many rare $K$ and $B$ decays. Even if they can be given a lower weight
  in the fit, they lower the precision and should be presently avoided.
\end{itemize}

\section{BV-Strategy}
In what follows I want to summarize the BV strategy developed in two papers
with Elena Venturini \cite{Buras:2021nns,Buras:2022wpw} which generalized my
2003 strategy used for $B_{s,d}\to\mu^+\mu^-$ decays \cite{Buras:2003td}
to all $K$ and $B$ decays. This strategy deals with the second and the third item
above but as I realized in \cite{Buras:2022qip} it  solves also the first
problem. It consists of five steps.

{\bf Step 1}

Remove CKM dependence from observables as much as possible by calculating suitable ratios of decay branching ratios  to
the mass differences $\Delta M_s$ and $\Delta M_d$ in the case of $B_s$ and
$B_d$ decays, respectively and to the parameter $|\varepsilon_K|$ in the case of Kaon decays. By suitable we mean for instance that in order to eliminate the $\vcb$
dependences in the branching ratios for $\kpn$ and $\klpn$, the parameter $|\varepsilon_K|$ 
has to be raised, as given later, to the power $0.82$ and $1.18$, respectively.
For $B_{s,d}$ decays one just divides the branching ratios by $\Delta M_{s,d}$, respectively.
In this manner CKM
dependence can be fully eliminated for all $B$ decay branching ratios. For
$K$ decays  only the dependence on the angle $\beta$ in the Unitarity Triangle (UT) remains.
The dependence on the angle $\gamma$ in the UT is practically absent so that future improvements
on the measurements of $\gamma$ by LHCb and Belle II collaborations will not
have any impact on these particular ratios although they will be very important
for other ratios discussed by us. On the other hand improved measurements of $\beta$ and improved values of hadronic parameters will reduce
the uncertainties in these ratios. It should be emphasized that already these
ratios constitute very good tests of the SM, simply because they are significantly more precise than the individual branching ratios.

{\bf Step 2}

Set $\Delta M_s$, $\Delta M_d$, $|\varepsilon_K|$ and the mixing induced
CP asymmetries  $S_{\psi K_S}$ and $S_{\psi \phi}$ to their experimental values.
This is done usually in global fits as well but here we confine the
fit to these observables. 
The justification for this step is the fact that within a good approximation  all these observables
can be simultaneously described within the SM without
any need for NP contributions \cite{Buras:2022wpw} and the theoretical and experimental status
of these $\Delta F=2$ observables is exeptionally good. In turn
this step not only avoids the tensions in the determinations of $\vcb$
and $\vub$ in tree level decays, but also provides in Step 5 SM predictions for numerous rare $K$ and $B$  branching ratios that are most accurate to date \cite{Buras:2022wpw,Buras:2021nns,Buras:2022qip}.

{\bf Step 3}

In order to be sure  that the $\Delta F=2$ archipelago is not infected
by NP a rapid test has to be performed with the help of the $\vcb-\gamma$ plot
\cite{Buras:2022wpw,Buras:2021nns}. This test turns out to be negative
dominantly thanks to the 2+1+1 HPQCD lattice  calculations of $B_{s,d}-\bar B_{s,d}$ hadronic matrix elements \cite{Dowdall:2019bea}\footnote{Similar results for $\Delta M_d$ and $\Delta M_s$ hadronic
    matrix elements have been obtained within the HQET sum rules in
    \cite{Kirk:2017juj} and \cite{King:2019lal,King:2019rvk}, respectively.}. The superiority of the
$\vub-\gamma$ plots over UT plots in this context has been emphasized in
\cite{Buras:2022nfn}. We will stress it again below.
\newpage
{\bf Step 4}  

As the previous step has lead to a negative rapid test we can now determine the
CKM parameters without NP infection on the basis of $\Delta F=2$ observables
alone. It should be noted that this step can be considered as a reduced
global fit of CKM parameters in which only $\Delta F=2$ observables have been taken into account.

{\bf Step 5}

All the problems listed above are avoided in this manner and having CKM
parameters at hand one can make rather precise SM predictions for the
observables outside  the $\Delta F=2$ archipelago and compare them with the
experimental data. In particular one can predict suitable $\vcb$-independent
ratios between various branching ratios not only within a given meson system
but also involving different meson systems. Such ratios depend generally
on $\beta$ and $\gamma$ with $\beta$ already precisely known and $\gamma$
determined in Step 4 more precisely than presently by the LHCb and Belle II
experiments.

In order to motivate this strategy in explicit terms, let us recall  the values of $\vcb$
extracted from inclusive and exclusive tree-level semi-leptonic $b\to c$ decays
\cite{Finauri:2023kte,FlavourLatticeAveragingGroupFLAG:2024oxs} 
\be
\vcb_\text{incl}=(41.97\pm0.48)\cdot 10^{-3},\qquad \vcb_\text{excl}=(39.21\pm0.62)\cdot 10^{-3}\,.
\ee
As rare K and B decays and mixing parameters are sensitive functions of $\vcb$, varying it from $39\cdot 10^{-3}$ to $42\cdot 10^{-3}$ changes $\Delta M_{s,d}$ and $B$-decay branching ratios
by roughly $16\%$, $\kpn$ branching ratio by $23\%$, $\varepsilon_K$ by $29\%$
and $\klpn$ and $K_S\to\mu^+\mu^-$ branching ratios by $35\%$.

These uncertainties are clearly a disaster for those like me, my collaborators
and other experts in NLO and NNLO calculations who spent decades to reduce theoretical uncertainties in basically all important rare $K$ and $B$ decays and
quark mixing observables down to $(1-2)\%$ \cite{Buras:2011we}. It is also a disaster for lattice QCD experts who for quark mixing observables and in particular  meson weak decay constants achieved the accuracy at the level of a few percents \cite{FlavourLatticeAveragingGroupFLAG:2024oxs}.

\section{The $\vcb$-Independent Ratios}\label{Ratios}

One constructs then in Step 1 a multitude of $\vcb$-independent ratios $R_i$ not only
of branching ratios to quark mixing observables but also of branching
ratios themselves. Those which involve branching ratios from different meson
systems depend generally on $\beta$ and $\gamma$. Once $\beta$ and $\gamma$
will be precisely measured, this multitude of $R_i(\beta,\gamma)$ will provide very good tests of the SM. However, using Step 5 of our strategy it is possible
to make SM predictions for these ratios already today and we will list several of them
below.

The details of the execution of this strategy  can be found in \cite{Buras:2022wpw,Buras:2021nns,Buras:2022qip}. In particular analytic expressions
for sixteen  ratios $R_i(\beta,\gamma)$ and plots for them can be found in 
\cite{Buras:2021nns}. A guide to these relations can be found in Section~4 of
the latter paper. 
Additional ratios,  predictions for all ratios considered  and  for 26 individual branching ratios resulting
from our strategy are presented in \cite{Buras:2022qip}.
Here we just list most interesting results obtained in these papers and update some of them.

We begin with  the most interesting $\vcb$-independent ratios that involve
both branching ratios and $\Delta F=2$ observables. They read
\be\label{F1}
  \boxed{\frac{\overline{\mathcal{B}}(B_s\to\mu^+\mu^-)}{\Delta M_s}=
    (2.13\pm0.07)\times 10^{-10}\text{ps}\,,}
\ee
   \be\label{F2}
\boxed{\frac{\mathcal{B}(B_d\to\mu^+\mu^-)}{\Delta M_d}= (2.02\pm0.08)\times 10^{-10}\text{ps}\,,}
\ee
  
\be\label{F3}
  \boxed{\frac{\mathcal{B}(\kpn)}{|\varepsilon_K|^{0.82}}=(1.31\pm0.05)\times 10^{-8}{\left(\frac{\sin\gamma}{\sin 64.6^\circ}\right)^{0.015}\left(\frac{\sin 22.2^\circ}{\sin \beta}\right)^{0.71},  }            }
  \ee
  \be\label{R12a}
\boxed{\frac{\mathcal{B}(\klpn)}{|\varepsilon_K|^{1.18}}=(3.87\pm0.06)\times 10^{-8}
    {\left(\frac{\sin\gamma}{\sin 64.6^\circ}\right)^{0.03}\left(\frac{\sin\beta}{\sin 22.2^\circ}\right)^{0.9{8}}.}}
\ee
The negligible dependence on $\gamma$ should be noticed so that the angle
$\beta$ plays more important role than $\gamma$ in this context. 
Using the experimental values of $\Delta M_s$, $\Delta M_d$, $|\varepsilon_K|$
and $\beta$ these ratios 
imply the most accurate predictions for the four branching ratios in question in  the SM  to date \cite{Buras:2022wpw,Buras:2021nns}. Moreover, they are independent of the value of $\vcb$. We will present them in Section~\ref{Magnificant}.

\begin{figure}
\centering
\includegraphics[width = 0.55\textwidth]{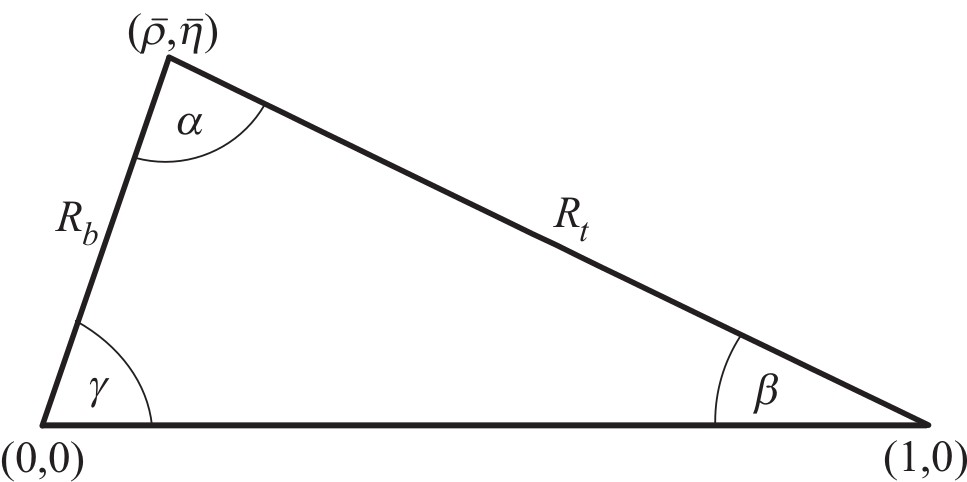}
 \caption{\it The Unitarity Triangle. }\label{UUTa}
\end{figure}

The ratios $R_i(\beta,\gamma)$ within the SM are given below.
The explicit expressions for these
ratios as functions of $\beta$ and $\gamma$ are given in \cite{Buras:2021nns}.  Here we just list the final results using the CKM parameters in (\ref{CKMoutput}) which were not given there. Moreover in the case of the ratios $R_5$ and
$R_7$ we use the most recent results for the formfactors entering $\mathcal{B}(B^+\to K^+\nu\bar\nu)$ from the HPQCD collaboration \cite{Parrott:2022zte,Parrott:2022rgu}. Here we go.
\be
\boxed{R_0(\beta)=\frac{\mathcal{B}(\kpn)}{\mathcal{B}(\klpn)^{0.7}}
={(2.03\pm 0.11)}\times 10^{-3}\,,
}
\label{eq:R0}
\ee
    \be\label{SR1}
\boxed{R_{\rm SL}=\frac{\mathcal{B}(\ksm)_{\rm SD}}{\mathcal{B}(\klpn)}=(6.29\pm0.52)\times10^{-3}\,,}
\ee

\be\label{R1}
\boxed{R_1(\beta,\gamma)=\frac{\mathcal{B}(\kpn)}{\left[{\overline{\mathcal{B}}}(B_s\to\mu^+\mu^-)\right]^{1.4}}= 53.23\pm2.75\,,}
  \ee
  \be \label{R2}
  \boxed{ R_2(\beta,\gamma)=\frac{\mathcal{B}(\kpn)}{\left[{\mathcal{B}}(B_d\to\mu^+\mu^-)\right]^{1.4}}= (8.37\pm0.47)\times 10^{3}\,,}
   \ee

   \be\label{R3}
   \boxed{R_3(\beta,\gamma)=\frac{\mathcal{B}(\klpn)}{\left[{\overline{\mathcal{B}}}(B_s\to\mu^+\mu^-)\right]^{2}}= (2.06\pm0.16)\times 10^6\,,}
   \ee
   \be\label{R4}
\boxed{R_4(\beta,\gamma)=\frac{\mathcal{B}(\klpn)}{\left[{\mathcal{B}}(B_d\to\mu^+\mu^-)\right]^{2}}=(2.83\pm0.24)\times 10^9\,,}
   \ee

\be\label{R5}
\boxed{R_5(\beta,\gamma)=\frac{\mathcal{B}(\kpn)}{\left[\mathcal{B}(B^+\to K^+\nu\bar\nu)\right]^{1.4}}={(2.32\pm0.13)\times 10^{-3}\,,}}
\ee
\be\label{R6}
\boxed{R_6(\beta,\gamma)=\frac{\mathcal{B}(\kpn)}{\left[\mathcal{B}(B^0\to K^{0*}\nu\bar\nu)\right]^{1.4}}=(8.43\pm1.21)\times 10^{-4}\,,}
\ee

\be\label{R7}
\boxed{R_7=\frac{\mathcal{B}(B^+\to K^+\nu\bar\nu)}{{\overline{\mathcal{B}}}(B_s\to\mu^+\mu^-)}={(1.30\pm0.07)\times 10^{3}\,,}}
\ee

\be\label{R8}
\boxed{R_8=\frac{\mathcal{B}(B^0\to K^{*0}\nu\bar\nu)}{{\overline{\mathcal{B}}}(B_s\to\mu^+\mu^-)}=(2.68\pm0.25)\times 10^{3}\,.}
\ee

It should be mentioned that to obtain these values we have used $\beta=22.2(7)^\circ$. A new average from  HFLAV~\cite{HFLAV:2022esi} reads 
$\beta=22.6(4)^\circ$ fully compatible with the one used by us but more
accurate. In view of other larger uncertainties it is better to wait for future
updates of $\beta$ from Belle II before all these ratios will be updated.
\section{New Look at the CKM Matrix and $\vcb-\gamma$ Plots}
Let us emphasize that our strategy  uses
as basic CKM parameters \cite{Blanke:2018cya,Buras:2022wpw,Buras:2021nns,Buras:2022qip}
\be\label{best}
\vus=\lambda, \qquad \vcb,\qquad  \beta,\qquad \gamma,
\ee
with $\beta$ and $\gamma$ being the two angels of the UT shown in Fig.~\ref{UUTa}.

This differs from the standard parametrization of the CKM matrix that
involves three real parameters and one complex phase
\be\label{LH}
s_{12}=\vus, \qquad   s_{13}=\vub,\qquad s_{23}=\vcb, \qquad \gamma,
\ee
 that can be determined  separately in tree-level decays.
Consequently,  basically all flavour phenomenology in the last three decades  used this set of parameters. In particular the determination of the UT
  was dominated by the measurements of its sides $R_b$ and $R_t$ through
  tree-level $B$ decays and the $\Delta M_d/\Delta M_s$ ratio, respectively, with   some participation of the measurements of the angle $\beta$ through the mixing  induced CP-asymmetries like $S_{\psi K_S}$, the parameter $\varepsilon_K$  and much less precise angle $\gamma$. This is the case not only
  of global analyses by UTfitter\cite{Bona:2007vi} and CKMfitter \cite{Charles:2004jd}  but also of less sophisticated
  determinations of the CKM matrix and of the UT.

  However, as pointed out in  \cite{Buras:2022wpw,Buras:2021nns,Buras:2022qip}, the most powerful strategy appears eventually to be the one which
  uses as basic CKM parameters the ones in (\ref{best}),
that is two mixing angles and two phases.
This choice is superior to the one in which $\beta$ is replaced by $\vub$ for
several reasons:
\begin{itemize}
\item
  The known tensions between exclusive and inclusive determinations of $\vcb$
  and $\vub$ \cite{Finauri:2023kte,FlavourLatticeAveragingGroupFLAG:2021npn}
  are represented only by $\vcb$ which can be eliminated efficiently
  by constructing suitable ratios of flavour observables $R_i(\beta,\gamma)$, see previous section,   which are free of the tensions in question.
\item
  As pointed out already in 2002 \cite{Buras:2002yj},
  the most
  efficient strategy for a precise determination of the apex of the UT, that is
  $(\bar\varrho,\bar\eta)$, is to use the measurements of the angles $\beta$ and $\gamma$. Indeed, among any pairs of two variables representing the sides and the angles of the UT that are assumed for this exercise to be known with the same precision, the measurement of $(\beta,\gamma)$ results in the most accurate values of $(\bar\varrho,\bar\eta)$. 
  The second best strategy would be the measurements of $R_b$ and $\gamma$. However, in view of the tensions between different determinations of $\vub$ and  $\vcb$, that enter $R_b$,  the $(\beta,\gamma)$ strategy
will  be a clear winner once LHCb and Belle II collaborations will improve  the measurements of these two angles.
  \item
  The $\vcb-\gamma$ plots for fixed $\beta$, proposed in  \cite{Buras:2022wpw,Buras:2021nns} are, as emphasized in \cite{Buras:2022nfn}, useful companions to
  common unitarity triangle fits because they exhibit better possible inconsistences between $\vcb$ and $(\beta,\gamma)$ determinations than the latter fits.
  We will demonstrate this below.
\end{itemize}

\begin{figure}[t!]
  \centering%
  \includegraphics[width=0.56\textwidth]{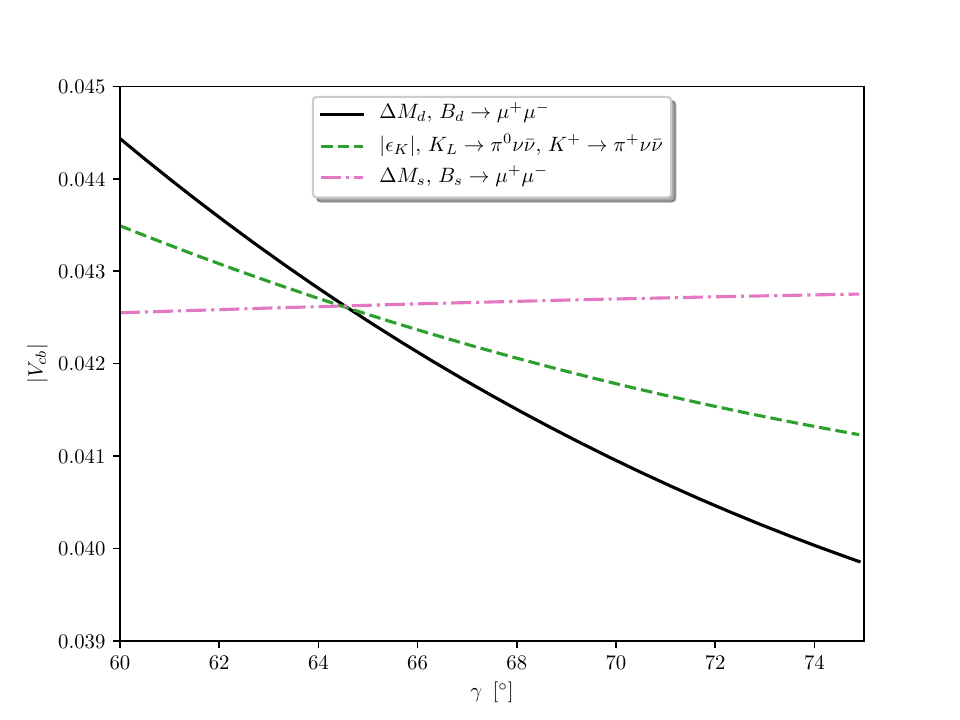}
  \includegraphics[width=0.42\textwidth]{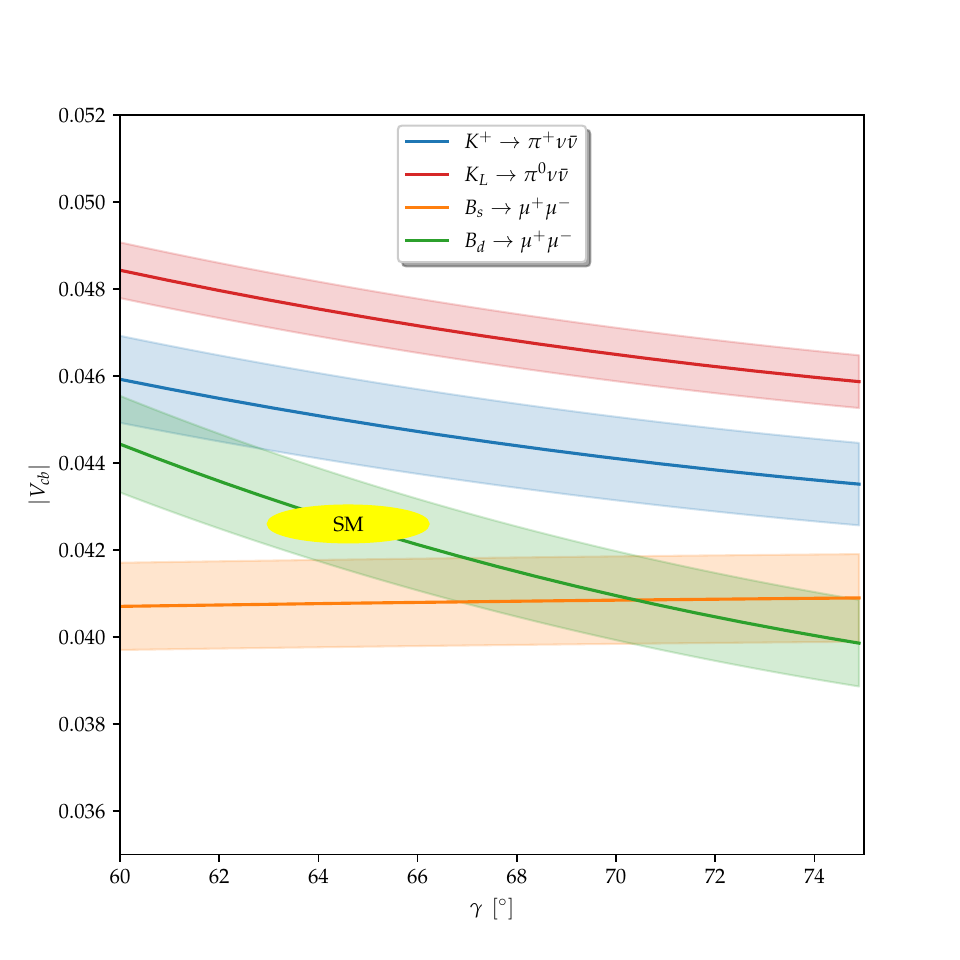}
  \caption{\it {Left:Schematic illustration of the action of the seven observables in the $\vcb-\gamma$ plane in the context of the SM. We set $\beta=22.2^\circ$ and all uncertainties to zero. Right:The impact of hypothetical future measurements of the branching ratios for $\kpn$, $\klpn$,  $B_d\to\mu^+\mu^-$ and $B_s\to\mu^+\mu^-$  on the $\vcb-\gamma$ plane. All uncertainties  are included. The yellow disc  represents the SM as obtained in {(\ref{CKMoutput})}.                                From \cite{Buras:2022nfn}.}
\label{fig:X}}
\end{figure}

In this context let us present two simple formulae that are central in the
$(\beta,\gamma)$ strategy as they allow to calculate the appex of the UT in no
time, but 
to my knowledge they have been presented  only recently for the first time  \cite{Buras:2023qaf}.
They read\footnote{This year one can celebrate the thirties birthday of
these parameters \cite{Buras:1994ec}, with their version without bars presented  earlier by Wolfenstein \cite{Wolfenstein:1983yz}.}
\be\label{AJB23}
\boxed{\bar\varrho=\frac{\sin\beta\cos\gamma}{\sin(\beta+\gamma)},\qquad
  \bar\eta=\frac{\sin\beta\sin\gamma}{\sin(\beta+\gamma)}.}
\ee
Evidently they can be derived by high-school students, but the UT is unknown to them and somehow no flavour physicist got the idea to present them in print until its first appearance in \cite{Buras:2023qaf}.

     The superiority of the  $\vcb-\gamma$ plots 
      with respect to $\vcb$ and  $\gamma$ over UT plots
      has been emphasized in  \cite{Buras:2022nfn}. Indeed,
\begin{itemize}
\item
  They exhibit $\vcb$ and its correlation with $\gamma$ determined through a given observable in the SM, allowing thereby monitoring the progress 
  on both parameters expected in the coming years. Violation of this correlation in experiment will clearly indicate NP at work.
\item
  They utilize the strong sensitivity of rare $K$ decay  processes to $\vcb$ thereby providing
  precise determination of $\vcb$
  even with modest experimental precision on their branching ratios.
\item
  They exhibit, as shown in Fig.~\ref{fig:X}, the action of $\Delta M_s$ and of $B_s$ decays,
like $B_s\to\mu^+\mu^-$
which is not possible in the common UT-plot.
\item
  Once the accuracy of $\gamma$ measurements  will approach $1^\circ$ it will be easier to monitor this progress on the $\vcb-\gamma$ plot than on the UT-plot.
\end{itemize}

In Fig.~\ref{fig:X}  we show examples of such plots that
can be used in the search for new animalcula. See figure caption for explanations.

\section{CKM Parameters}
As the rapid test for the $\Delta F=2$ observables turned out to be negative we can now determine the CKM parameters using these observables without 
NP infection. We find \cite{Buras:2022wpw}
\be\label{CKMoutput}
\boxed{\vcb=42.6(4)\times 10^{-3}, \quad 
\gamma=64.6(16)^\circ, \quad \beta=22.2(7)^\circ, \quad \vub=3.72(11)\times 10^{-3}\,}
\ee
and consequently
\be\label{CKMoutput2}
\boxed{\vts=41.9(4)\times 10^{-3}, \qquad \vtd=8.66(14)\times 10^{-3}\,,\qquad
{\IM}\lambda_t=1.43(5)\times 10^{-4}\,,}
\ee
\be\label{CKMoutput3}
\boxed{\bar\varrho=0.164(12),\qquad \bar\eta=0.341(11)\,,}
\ee
where $\lambda_t=V_{ts}^*V_{td}$. Note that our result for $\gamma$ agrees perfectly with the most recent LHCb measurement: $\gamma=64.6(28)^\circ$.

\begin{table}
\centering
\renewcommand{\arraystretch}{1.4}
\resizebox{\columnwidth}{!}{
\begin{tabular}{|ll|l|}
\hline
Decay 
& SM Branching Ratio
& Data
\\
\hline \hline
 $B_s\to\mu^+\mu^-$ &  $(3.78^{+ 0.15}_{-0.10})\cdot 10^{-9}$      &  $(3.45\pm0.29)\cdot 10^{-9}$ \cite{HFLAV:2022pwe} 
\\
 $B_d\to\mu^+\mu^-$ &  ${(1.02^{+ 0.05}_{-0.03})}\ \cdot 10^{-10}$      & $\le 2.05\cdot 10^{-10}$ \cite{LHCb:2021awg}
\\
$B^+\to K^+\nu\bar\nu$ & $(4.92\pm 0.30)\cdot 10^{-6}$
&    $ (13\pm 4)\cdot 10^{-6}$ \cite{Belle-II:2023esi}
\\
$B^0\to K^{0*}\nu\bar\nu$ & ${(10.13\pm 0.92)}\cdot 10^{-6}$ &
 $\le 1.5\cdot 10^{-5}$ \cite{Grygier:2017tzo}
\\
\hline
$\kpn$ & $(8.60\pm 0.42)\cdot 10^{-11}$ &  $(13.0\pm 3.1)\cdot 10^{-11}$ \cite{NA62:2022hqi,NA62:2024pjp} 
\\
 $\klpn$ & $(2.94\pm 0.15)\cdot 10^{-11}$ &   $\le 2.0\cdot 10^{-9}$ \cite{Ahn:2018mvc} 
\\
$(\ksm)_{\rm SD}$& {$(1.85\pm 0.12)\cdot 10^{-13}$} &   $\le 2.1\cdot 10^{-10}$
\cite{LHCb:2020ycd}
\\
\hline
\end{tabular}
}
\renewcommand{\arraystretch}{1.0}
\caption{\label{tab:SMBRs}
  \small
  Present most accurate  SM estimates  of the branching ratios for the {\em the  magnificant seven} 
  obtained using the BV-strategy.
  }
\end{table}

\section{The Magnificant Seven}\label{Magnificant}
Until now we dealt only  with the issue of the CKM parameters but the choice
of the observables that are particularly theoretically clean is also
very important. In my view the seven ones listed in Table~\ref{tab:SMBRs} are particulary
promising. We give there their SM values obtained using our strategy together
with the present experimental results.
Clearly other decays, in particular
$B\to K\ell^+\ell^-$ and $B\to K^*\ell^+\ell^-$, that remain still central in the
search for NP, are very important. Unfortunately, there are different views
among theorists on the role of long distance QCD effects in these decays
and the interpretation of the observed anomalies in these decays is difficult
at present. It should be noted that in the case of $\ksm$ also long distance
contributions are present but fortunately as demonstated in
\cite{Dery:2022yqc}, generalizing
the ideas of \cite{DAmbrosio:2017klp}, the short distance contribution can be extracted and measured.

The strong suppression of NP in $\Delta F=2$ prosesses  imposes stringent constraints on the parameter space of $Z^\prime$ models and limit their potential effects in rare decays. However, as emphasized in \cite{Buras:2014sba,Buras:2014zga,Crivellin:2015era}, the NP contributions to
$\Delta F=2$ processes in these models can be suppressed for a particular pattern of left-handed and right-handed flavour-violating $Z^\prime$ couplings
so that NP effects in rare K and B decays can be sizable.

\begin{figure}[t]
  \begin{center}
    \includegraphics[width=0.42\textwidth]{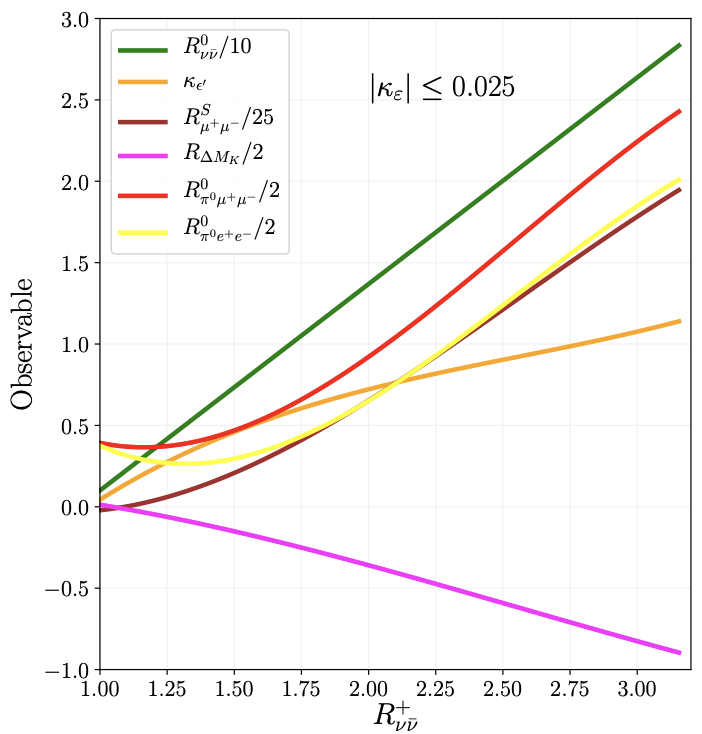}
    \includegraphics[width=0.56\textwidth]{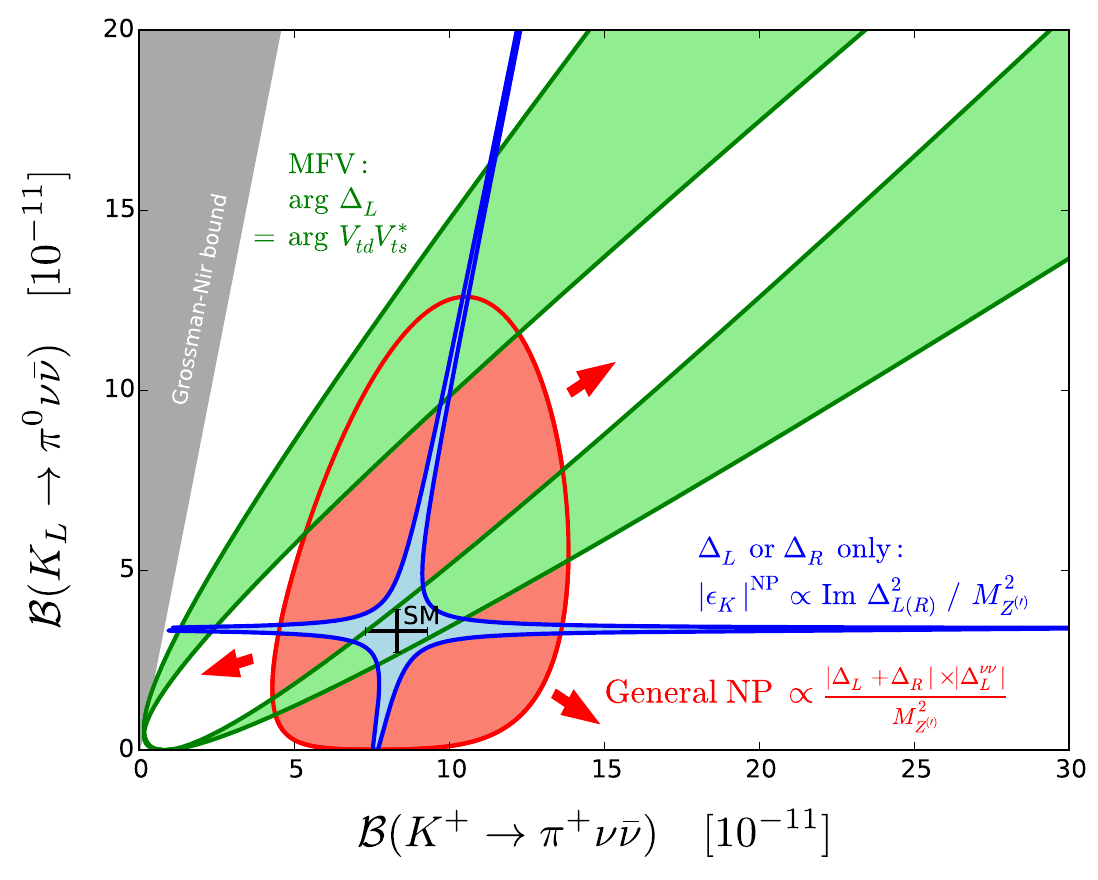}
\caption{Left:Correlations between the observable 
  $R_{\nu \bar \nu}^+$ and various other Kaon observables in a $Z^\prime$ model.
  All ratios $R_i=1$ in the SM. Note that for $\klpn$ and $K_S\to\mu^+\mu^-$ the
  ratios are divided by 10 and 25 respectively.  From \cite{Aebischer:2023mbz}.
  Right:Correlations between $\kpn$ and $\klpn$ in general $Z^\prime$ models with
  green region representing MFV models and blue one coming from $\varepsilon_K$
constraint with only LH or RH couplings \cite{Blanke:2009pq}. From \cite{Buras:2015yca}.}
\label{Fig:3}
\end{center}
\end{figure}

The implications of the suppression of NP contributions to $\varepsilon_K$ on
rare K decays 
within $Z^\prime$ models have been studied in \cite{Aebischer:2023mbz}. In this case 
it is sufficient to choose the relevant left-handed couplings to be purely imaginary implying strong correlations
between three Kaon decays listed in Table~\ref{tab:SMBRs} and the ratio $\epe$ as seen in the left plot in Fig.~\ref{Fig:3}. The strict correlation between $\kpn$ and $\klpn$
decays in the presence of strong suppression of NP to $\varepsilon_K$ seen in this figure is an example of a general correlation between
these decays pointed out by Monika Blanke long time ago \cite{Blanke:2009pq}.
Its version in $Z^\prime$ models is presented in the right plot 
 in Fig.~\ref{Fig:3} taken from \cite{Buras:2015yca}.

In this context the recent NA62 result for $\kpn$,
listed in Table~\ref{tab:SMBRs}, implies
\be
\boxed{R^+_{\nu\bar\nu} = 1.51\pm 0.35\,\qquad (\text{SM}:~R^+_{\nu\bar\nu} = 1).}
\ee
making hopes that the branching ratios for $\klpn$ and $K_S\to\mu^+\mu^-$
could be strongly enhanced over the SM values. However, to this
end the experimental error on $\kpn$ branching ratio should be decreased.

In the case of B decays sufficient suppresion of NP contributions to $\Delta M_s$ and $\Delta M_d$ requires the presence of both left-handed and right-handed $Z^\prime$ couplings to quarks. The corresponding analysis for rare
B decays has been performed in \cite{Buras:2024mnq} finding again strong correlations between
B-decay branching ratios in Table~\ref{tab:SMBRs} and the ones for
$B\to K\ell^+\ell^-$ and $B\to K^*\ell^+\ell^-$ allowing for the explanation
of the anomalies in the latter ones despite the suppression of NP contributions
to $\Delta F=2$ processes\footnote{ 
In 331 models, in which only left-handed $Z^\prime$
couplings are involved, some amount of NP contributions at the level of
$5\%$ has to be allowed to explain the anomalies in question \cite{Buras:2023ldz}.}. The most interesting result of the analysis in \cite{Buras:2024mnq} is the
finding that in this scenario the  suppression of $B\to K(K^*)\mu^+\mu^-$ 
      branching ratios observed by the LHCb implies {\em uniquely}  enhancements of $B\to K(K^*)\nu\bar\nu$ branching ratios as hinted by the most recent Belle II data.

In the case of decays with neutrinos in the final state the SM predictions are obtained assuming the Dirac nature of neutrinos. However they could be of Majorana type
implying different predictions. Moreover, as neutrinos are invisible, what
is really measured are the decays  $K\to\pi+E_{\rm miss}$ and $B\to K(K^*)+E_{\rm miss}$ with $E_{\rm miss}$ standing for the missing energy. This means
that instead of neutrinos there could be different neutral particles like scalars,
fermions and vectors in the final state including dark particles. Strategies
for disentangling these possibilities  through kinematic distributions in the missing energy $E_{\rm miss}$  have been developed recently in \cite{Buras:2024ewl}.
See also earlier papers listed there. In particular a recent paper on the anatomy of such distributions in $\kpn$ \cite{Gorbahn:2023juq}. In fact such a distinction is possible although in many cases it will be a challenge for experimentalists.

\section{Summary and Shopping List}\label{outlook}\label{sec:7}
Despite the presence of various anomalies in the existing data, it is
clearly not evident which animalcula could be responsible for them. Possibly,
the main candidates are $Z^\prime$ vector bosons, leptoquarks and vector-like
quarks and vector-like leptons but to find out without any doubts what they are  we need more data,
in particular for theoretically clean decays as the ones listed in Table~\ref{tab:SMBRs}. Precise measurements of branching ratios for these decays and of
various kinematical distributions should allow in this decade to find
out which animalcula are responsible for them. The strategies presented
here should be helpful in this respect, in particular the ratios
$R_i(\beta,\gamma)$ listed in Section~\ref{Ratios}
independenty of whether NP effects are strongly suppressed in  $\Delta F=2$ processes or not.

Let me finish this writing with my shopping list for the coming years.
I list here only the entries related 
to flavour physics.
\begin{enumerate}
\item
  Precise measurements of the branching ratios for {\em the  magnificant seven} and
  of missing energy distributions for four decays among them.
Presently on a forefront are  the $B\to K^{(*)}\nu\bar\nu$ decays studied intensively by the Belle~II experiment~\cite{Belle-II:2023esi}, giving some
hints for NP contributions. For selected recent analyses of these data see
\cite{Buras:2024mnq,Bause:2021cna,He:2021yoz,Bause:2022rrs,Becirevic:2023aov,Bause:2023mfe,Allwicher:2023xba,Dreiner:2023cms, Altmannshofer:2023hkn, Gabrielli:2024wys, Hou:2024vyw, He:2024iju, Bolton:2024egx, Marzocca:2024hua,Allwicher:2024ncl}.
The same applies to $\kpn$ for which a very interesting result has 
been provided very recently by the NA62 experiment listed in Table~\ref{tab:SMBRs}.
Taking the present experimental result for $B_s\to \mu^+\mu^-$ allows us
to determine the present $\vcb$-independent experimental values for the
the ratios $R_1$ and $R_7$ in (\ref{R1}) and (\ref{R7})
that read
\be
\boxed{R_1=91.4\pm22.8,\,  (\text{SM}:53.23\pm2.75)}
\boxed{R_7=(3.8\pm1.2)\cdot 10^{-3}, \,
  (\text{SM}:(1.30\pm0.07)\cdot 10^{3}).}
\ee
They differ significantly from very precise SM values. Let us hope that the experimental errors on these ratios will decrease in the coming years.
\item
  Precise measurements of the branching ratios and of other observables in $B\to K\ell^+\ell^-$ and $B\to K^*\ell^+\ell^-$ decays and the clarification of the
  anomalies in them.
\item
  Precise measurements of $\vcb$ and $\gamma$ in tree-level decays that
  would provide additional tests of the BV-strategy.
 \item
   Clarification of anomalies in $B\to\pi K$ decays \cite{Buras:2003dj,Buras:2004ub,Fleischer:2018bld,Berthiaume:2023kmp,Datta:2024zrl,Szabelski:2024cem,Altmannshofer:2024kxb}.
 \item
   The $\Delta I=1/2$ rule for $K\to\pi\pi$ is almost 70 years old and we do not
   yet know whether it is fully explained by the SM.
The basic QCD
dynamics behind this rule - contained in the hadronic matrix elements
of current-current operators - has been identified analytically first in 1986 in the framework of the Dual QCD (DQCD)  \cite{Bardeen:1986vz} with some improvements
in 2014  \cite{Buras:2014maa}. This has been  confirmed more than 30 years later
by the RBC-UKQCD collaboration \cite{RBC:2020kdj} although the modest accuracy of both approaches still allows for some NP contributions. See
\cite{Buras:2022cyc} for the most recent summary. It would be important
to have two lattice calculations with $2+1+1$ flavours.
\item
  Similar comments apply to $\epe$ for which the only Lattice QCD calculation
  is from  the RBC-UKQCD with $2+1$ flavours and without isospin breaking effects
   \cite{RBC:2020kdj}.
  Hopefully it will be improved in the coming years and also performed by another QCD lattice collaboration.
A stressed in  \cite{Buras:2023qaf}
   Jean-Marc G{\'e}rard and myself expect on the basis of DQCD developed
   with Bill Bardeen \cite{Bardeen:1986vp,Bardeen:1986uz,Bardeen:1986vz}, significant NP contributions to $\epe$ that as seen
   in the left plot in  Fig.~\ref{Fig:3} are consistent with the
   recent result on $\kpn$  of the NA62 collaboration.
 \item
   After 37 years of waiting \cite{Bardeen:1987vg,Gerard:2010jt,Buras:2014maa}
   Lattice QCD 
   confirmed the DQCD claim that the $\hat B_K\le 0.75$ with the most accurate
   result from the RBC-UKQCD collaboration 
   $\hat B_K=0.7436(82)$ \cite{Boyle:2024gge} to be compared with
   $\hat B_K=0.73\pm0.02$ from DQCD \cite{Buras:2014maa}. As the ETM collaboration found  $\hat B_K=0.717(18)(16)$ \cite{Carrasco:2015pra} it is likely that
   the central value of $\hat B_K$ is very close to the DQCD one. This would
   imply that if there is some small NP contribution to $\varepsilon_K$ it
   must be of the same sign as the SM one. Let us hope that further
   improvements on $\hat B_K$ and NP matrix elements
   \cite{Boyle:2024gge,Buras:2018lgu,Carrasco:2015pra,Jang:2015sla} will be made in the coming years.\\
   The above statement has been made in the first version of our paper. Recently, a NNLO calculation for $\hat B_K$ has been performed in \cite{Gorbahn:2024qpe} which decreased the error in the RBC-UKQCD result but moved it upwards:
   $\hat B_K=0.7600(53)$. Thus the story of $\hat B_K$ is not finished yet.
   However, what is quite amazing is the fact that  $\hat B_K$ is so close to
   its strict large $N$ limit: $\hat B_K=0.75$. Presently, to my knowledge, only
   DQCD can explain the dynamics behind this result. It is the strong cancellation of the negative pseudoscalar meson contributions to  $\hat B_K$ by the positive vector meson contributions in the context of the so-called meson evolution within DQCD
   \cite{Buras:2014maa}.
 \item
   Calculation of hadronic matrix elements relevant for $\Delta M_s$ and
   $\Delta M_d$ with  $2+1+1$ flavours by other lattice QCD collaborations in order to check HPQCD results
   for these matrix elements.
 \item
   Further search for lepton flavour violation.
   \end{enumerate}

We should have then great time in the rest of this decade but unfortunately this does not depend entirely on flavour physicists as we have recently seen in
the case of CERN decision on the HIKE experiment. In view of the recent
NA62 result on $\kpn$ this could turn out to be  a big mistake.

%

\section*{Acknowledgements}
 It is 
a pleasure to thank the organizers of this workshop for inviting me 
to this interesting event and for an impressive hospitality. Particular thanks 
go to Elena Venturini  for fantastic time we spent together developing our
strategy. Thanks go also to Jason Aebischer, Jacky Kumar and Peter Stangl
for collaborations on the implications of the suppressed NP contributions to
$\Delta F=2$ processes and to Julia Harz and Martin Mojahed for developing
strategies for disentangling NP in $K\to\pi\nu\bar\nu$ and $B\to K(K^*)\nu\bar\nu$ decays.  Financial support from the Excellence Cluster ORIGINS,
funded by the Deutsche Forschungsgemeinschaft (DFG, German Research
Foundation), Excellence Strategy, EXC-2094, 390783311 is acknowledged.

\bibstyle{woc}
\bibliography{Bookallrefs}
\end{document}